# Theory of Transmission of Light by Sub-wavelength Cylindrical Holes in Metallic Films

N. García and Ming Bai

Laboratorio de Física de Sistemas Pequeños y Nanotecnología,

Consejo Superior de Investigaciones Científicas, Serrano 144 Madrid 28006, Spain

**Abstract**

This paper presents theory and finite-difference time-domain (FDTD) calculations for a single and arrays of sub-wavelength cylindrical holes in metallic films presenting large transmission. These calculations are in excellent agreement with experimental measurements. This effect has to be understood in terms of the properties exhibited by the dielectric constant of metals which cannot be treated as ideal metals for the purpose of transmission and diffraction of light. We discuss the cases of well-differentiated metals silver and tungsten. It is found that the effect of surface plasmons or other surface wave excitations due to a periodical set of holes or other roughness at the surface is marginal. The effect can enhance but also can depress the transmission of the arrays as shown by theory and experiments. The peak structure observed in experiments is a consequence of the interference of the wavefronts transmitted by each hole and is determined by the surface array period independently of the material. Without large transmission through a single hole there is no large transmission through the array. We found that in the case of Ag which at the discussed frequencies is a metal there are cylindrical plasmons at the wall of the hole, as reported by Economu et al 30 years ago, that enhanced the transmission. But it turns out, as will be explained, that for the case of W which behaves as a dielectric, there is also a large transmission when compared with that of an ideal metal waveguide. To deal with this problem one has to use the measured dielectric function of the metals. We discuss thoroughly all these cases and compare with the data.

**PACs Code**: 42.25.Bs, 42.79.Gn

# I. INTRODUCTION

Experiments were reported [1] showing that the transmission of light through sub-wavelength holes drilled periodically in a metallic film of Ag was large, 1000´s times larger, as compared with the transmission of one single hole of the same size in the same material. Recent experiments [2], by part of the same team of ref. 1(Lezec and Thio), appear to contradict the earlier experiments [1]. The explanation of the experiments [1] where based on the existence of surface plasmons polaritons (SPP) that are excited in the case of a set of periodic holes. After the initial work, a whole serial of papers have appeared insisting on the same point (for a review see ref.2). The last one, to our knowledge, appears recently on the same matter [3]. The recent paper by Lezec and Thio [2], reviewing the field and containing many new data, criticised much the whole saga of papers on the matter of the extraordinary enhancement of the transmission in a periodic array of holes in metallic films. This paper [2] claims to dismount the interpretation of SPP to understand the experiments and disclaims the entire picture in periodic arrays of holes. Also, it showed experimentally that the transmission enhancement by the periodic array with respect to that of a single hole is as much as a factor of 7, not a factor of 1000s, so that it can be a depression of the relative transmission as well. In fact the title of this paper is: *"Diffracted evanescent wave model for enhanced and suppressed optical transmission through subwavelengths hole arrays"*. Maybe there is a point of broken physical argument in ref. 1 and subsequent papers. Their claims are based on that they compared the transmission by each hole of the array with the transmission of a single hole reported in an earlier paper by Bethe [4]. This paper [4] was a theoretical study showing that the transmission of a sub-wavelength hole drilled in a perfect metal screen, ideal conductor (dielectric constant being negative infinity, $\varepsilon \to -\infty$), behaves as $(r/\lambda)^4$, where $\lambda$ is the wavelength of the radiation and $r$ is the hole radius. However ref.4 makes an approximation that is only valid for holes much smaller than $\lambda$, so that the field is practically constant on the hole. This does not hold for the experiments reported [1-3] because for the frequencies considered, the hole radius should be smaller than 25 nm but the holes used in the experiments are much larger. Another point is that an ideal metal has little resemblance regarding optical propagation with plasmonic metals as Ag or with a dielectric as W for the experimental frequencies. This has been discussed in part in a recent work [5]. Therefore the Bethe paper [4] has no significance in the problem at hand. In fact the main result in ref.2, in our opinion, is that the single whole transmission is very large when compared with that reported in ref.4 or with that resulting from a theory for wave propagation in ideal metals long waveguides. Therefore it seems reasonable to try to discuss and understand these experiments using the experimental frequency dependent dielectric constants for the metals Ag and W. For these two materials we used extensive comparisons with the data.

The aim of this paper is to study the transmissions of a hole of sub-wavelength size in a flat metallic film with thickness of the order of the hole diameter, as those used experimentally [1-3]. When simplistic estimations are considered, the transmission can be of the order of several thousand times larger than that for the same hole in an ideal metal. The understanding of this phenomenon is due to two terms for Ag: i) surface plasmons excited at the cylinder walls defining the hole, having the same nature that those described by Pfeiffer, Economou and Ngai [6] also Martinos and Economou [7] for metallic cylinders and recently discussed in the contest of the problem at hand [5]. These are surface plasmons rotating at the surface of the cylinders and propagating along its axis. ii) the penetration of the field in the metal. For the dielectric metal W, there may be a question whether cylindrical plasmons exist, but still is the penetration of the field in the metal. The signatures of waves at the cylinder surface are well manifested in the intensity curve depending of frequency for single holes. As will be seen these are effects that are governed in a subtle way by the values of the dielectric function at each frequency.

## II. CALCULATIONS

The structures we calculated in this work are single circular hole or hole arrays in a flat metallic film with a given diameter *d* and thickness *t*. Fig.1 a and b present a view of the single and arrays of hole structures with period P. The plane wave impinges perpendicular to the plane of the holes. plane pulse wave with a broad band is set as incident wave. We recorded both the incident and transmitted wave through the structure. The frequency response of the structure is calculated by dividing the spectrum of the transmitted wave by the spectrum of the incident wave.

The calculations are performed using 3D-FDTD method. The wave spectrums before and after the structures are achieved by Fourier transform from the recorded wave signal in time domain.

In the FDTD method we used for different materials, the metals like Ag and W are modelled as frequency dispersive media. The corresponding frequency dependent permittivities are retrieved from experimental data by Johnson and Christy [8] and from Physics Data [9]. For ideal metal, it is treated by means that the electric components of wave are excluded from any part of the metal; i.e. the field is forced to be zero at the metal surface. This is equivalent to set the permittivity as infinitely negative.

The perfect matched layer absorbing boundary condition [10] is applied in FDTD calculations for single hole transmission, and periodical boundary condition are applied for hole arrays transmission. The grids and time steps are taken fine enough to obtain convergent solutions.

**II. A. The Ag Case for Single Holes**

*i) Calculations*

Ag is a paradigmatic case for studying surface plasmons and there are a large amount of literatures on surface polariton plasmons (SPP) by gratings [11] in the visible region. The reason for that is because at these wavelengths the imaginary part of the permittivity ($\varepsilon_2$) is small, then the SPP is well defined [11, 12]. Fig.2a shows the real ($\varepsilon_1$) and the imaginary ($\varepsilon_2$) part of the permittivity. It represents a nice Drude plasmonic behaviour with a bulk plasmon wavelength, $\lambda_p$=325nm ($\omega_p$~3.8ev). It has been also shown that in a periodic surface with a certain small single Fourier component the enhancement of the field due to the SPP can be very large ($\approx$ 100 times that of the incident field) [12,13]. However when the Fourier component increases or many Fourier component exist (for example a step-like profile of the surface) the enhancement is reduced drastically due to the enlargement of the SPP linewidth. We mention these points to stress that the existence of SPP in a surface does not imply large enhancements but many other requisites are needed. In particular the structure for study in Fig.1b has many Fourier components and the SPP enhancement due to SPP can not be large.

To model metal Ag in the calculation, a Drude dispersion relation (Equ. 1) is used to meet the frequency dependent permittivity of Ag.

$$\varepsilon(\omega)/\varepsilon_f = 1 - \frac{\omega_p^2}{\omega^2 - 2i\omega\delta} \qquad (1)$$

which is characterized by fitting permittivity $\varepsilon_f$ in the visible range, the bulk plasma frequency $\omega_p$ and the damping constant $\delta$. These parameters are chosen to fit the experimental data by Johnson and Christy [8] for Ag. The chosen parameters are $\varepsilon_f$=6.8, $\omega_p$~3.8eV and $\delta$~-0.02eV. In the frequency region that we are interested, the whole visual range, these values give a satisfactory good fit with the experimental data for both the real and the imaginary parts of the permittivity, as shown in Fig 2a.

The transmission coefficient is presented in Fig.2b for a hole of diameter $d$=270nm in metal film with thickness $t$=340nm for the permittivity (Eq.1). The incidence is in normal direction with respect to the surface of the metal film. We also present results for a hole in ideal metal film with the same diameter and with thickness of $t$=340nm and 750nm. The results are quite illuminating and tell us what is going on in the single hole transmission. It is clear that the transport using the theory of transport in waveguides in ideal metals which shows the first wavelength cutoff at $\lambda_c$=2$\pi d$/3.68=1.705$d$ [14] only applies for large values of $t$=750nm (see Fig.2b). For comparison, the waveguide theory for ideal metal with much larger thickness is also plotted. For the values of $t$ smaller than 340nm which is used in the experiments, a hole in the ideal metal gives a considerable transmission in long wavelength tail. Even at $\lambda$=700nm, the coefficient is 0.10. Therefore these result show that making considerations for enhancement of transmission comparing with long waveguide in ideal metals is unphysical and not realistic.

The striking result is that when calculate the transmission with Ag film using the experimental values of permittivity for Ag, we found a much higher transmittivity at larger wavelengths. The cutoff has been moved from 460nm (ideal waveguide cutoff for $d$=270nm) to ~630nm. Notice the results presented in Fig.5a, at 700nm where a transmittivity peak of 1.4 appears for the hole arrays, one single hole provides a transmittivity of 0.5 in Fig 2b. So the transmission enhancement from hole arrays versus single hole is only a factor of 2.8. Then the question is: *why has the real Ag the behaviour of large transmission at large wavelengths?* This has to be searched in the effect of surface modes located at the cylindrical cavity defining the holes. They are similar to those discussed by Pfeiffer, Economou and Ngai [6] also Martinos and Economou [7] for metallic cylinders. These waves move the cutoff of transport in the structure to longer wavelengths and are the responsibility of the large transmissions through the hole arrays for the observed peak at 700nm for the $t$=340nm as is the case of the experiments [1,3].

*ii) Cylindrical surface waves.*

The surface plasmons excited in the cylindrical holes propagate the same way as those that has been studied on the surface of metallic cylinder [6, 7]. The surface plasmons locate along the circumference of the cylinder with wavelength $\lambda^{\theta}_n=2\pi r/n$, i.e. circumference length divided by the index branch $n$. And the SPs also propagate along the cylinder axis z with a wave $k_z$. The possibility of exciting long wavelength modes is given by the cylindricality $\alpha=2\pi r/\lambda_p$, where $\lambda_p$ is the bulk plasmon wavelength ( $\lambda_p$=325nm for Ag [8] ). This theory is done for Drude plasmon dispersion relation with $\delta$=0 in (1). In our case we do not have a cylinder of infinite length as has been discussed in ref. 5 and 6, but we have a cylindrical metallic cavity of certain thickness $t$. However by looking at the boundary conditions the same kind of modes should exist and our full solution of Maxwell equations shows up in the transmission. In Fig. 4a modulations in the transmission can be identified, which actually may correspond to the surface plasmons modes excited in the cavity surface. The peak with longer wavelength corresponds to the smaller $n$, $n$=1 identifies the longest wavelength surface plasmons mode, i.e. there is a cutoff for the surface plasmons modes.

This cutoff of the surface plasmons is determined by the dispersion relation of the n=1 mode. The dispersion relation was studied with a planar approximation in ref. 6. As an approximation the cylindrical surface was treated as semi-infinite plane, the curvature of the cylinder was considered as periodic boundary condition. The resulted dispersion relation in the surface can be written as [6]:

$$\omega_{sp} = \omega_p [1 + \frac{1}{2}Q^{-2} + (1+\frac{1}{4}Q^{-4})^{\frac{1}{2}}]^{-\frac{1}{2}}$$
$$Q = [K^2 + (n/\alpha)^2]^{\frac{1}{2}} \quad K = k/k_p$$

(2)

where $k_p = 2\pi/\lambda_p$ and the cylindricality $\alpha=dk_p/2$ *play an important role in the dispersion relation.*. Fig.3a actually is the case for cylinder holes in Ag film with $\lambda_p$=325nm, diameter $d$=270nm, the same as in Fig 2b As shown in Fig. 3a, with fixed cylindricality parameter $\alpha$, the possible modes of surface plasmons that can be excited at the cross points with the photon line are limited in wavelength.. From Fig.3a, the wavelength for the possible surface plasmons can be excited in the cylinder holes is in range of 470nm~630nm. The 630nm corresponds approximately with the cutoff for the transmission using the permittivity in Fig.2b. There are also weak oscillations in the structure of the transmission, in both experiments and calculations that we may tentatively assign to the different plasmons index *n*, in Fig.3a and in Fig.4 as well.

However, we noticed that the above dispersion relation is based on the semi-infinite plane approximation for the cylindrical surface for a Drude metal with δ=0. The peak structures are not expected to be completely matched with the experiments. In fact the *n* index in our cylindrical hole structure may depart considerably of those given by (2) and plotted in Fig.3a to illustrate the problem. Still our holes are of finite small thickness while the theory of ref.6 is for infinite cylinders. The more important point is that this theory result explains the extra transmission above the waveguide cutoff limit. The cylindrical wave is excited at the entrance of the hole and it carries the energy through the other side of the hole which is not allowed in ideal waveguide. It is assuming that the thickness of the hole is not big enough because the wave has a propagation decay length. When the thickness of the hole becomes larger than the decay length, the transmission will be controlled by the waveguide modes without cylindrical surface waves. We have performed calculations for Ag with $d$=270nm and changing the thickness $t$=340nm, 525nm and 735nm to check the propagation decay length, as shown in Fig. 3b. Clearly when $t$=735nm, the cutoff is retracted to 600nm, however still bigger than that of the ideal metal cutoff (~500nm) for $t$=750nm in Fig.2b. This establishes that the transmission in a hole of Ag at larger wavelengths (>500nm) is controlled by cylindrical surface waves with a decay length that we estimated to be more than 1um, which we will discuss in anther work. The decay should depend also of the diameter because it limits the extension of the cylindrical wave into the vacuum in the hole, as well as of the thickness.

*iii) Comparison with Experiments*

Ref.2 has presented an ample number of experiments for single holes of different values of *d* and *t*. These represent a good set of experimental data to contrast with our calculations. Fig.4a presents the experimental data (Ref.2, Fig.2c) as well as the calculations for the same parameters as in the experiments. The comparison is strikingly good for all the cases. Also, as is important we plot the enhancement for the case of *d*=270nm. This is defined as the transmission for the real Ag divided by the transmission predicted by the ideal waveguide theory [14]. It is observed that this enhancement can be up to 1000. Analogously we present similar results in Fig.4b for the case of *d*=200nm and the enhancement is of the order of 10000. This proves that the enhancements by single holes we calculated are already of the order of those measured for hole arrays in Ref.1 and claimed to be due to SPPs of the periodical arrays. Calculations and experiments [2] prove that it is not necessary to have hole arrays in order to have such enhancements. This tends to rule out the SPP between the arrays as the physical reason for enhancement from arrays. However it is due to the cylindrical surface waves in the walls of the cavity drilled on the metal to have enhancement from single hole. It is clear that the influence of SPPs, if exist, is marginal for the large values observed in the transmission from holes arrays. It is also worth to notice that the transmission of a single hole shows weak oscillations as we tentative assign to the different cylindrical surface waves index *n* indicated in Fig.3a.

**II.B. The Ag Case for Arrays of Holes**

We now proceed by discussing the transmission of light through an array of holes following the same procedures discussed above. Since a single hole gives such an enhancement beyond the cut-off wavelength, it will not be a surprise that a periodical array will give also a very large enhancement. The result for an array will be produced by the interference of the waves merging from the holes. Therefore the transmission will have for some frequencies enhancements over the single hole transmission and for other frequencies depressions. Same ideas have been described in ref. 2, however our FDTD calculations will prove all at once. In order to prove this we have performed such calculations for periodical arrays of holes to compare with existing experimental results [2, 3]. Fig.5a shows the transmission results for the periodical array with P=600nm,

$d$=270nm and $t$=225nm, the same parameters corresponding to the data presented in Fig.1 of ref.3. The agreement is again strikingly remarkable and without fitting any parameter, just taking the permittivity of Ag [8]. The three experimental peaks at $\lambda \approx$ 700nm, 550nm and 430nm are excellently described not only the peaks positions but also the measured amplitude. For comparison we also presented the existing theory performed in ref. 3 in which a rather good agreement is claimed. In fact, there is not such an agreement because the calculations only show two peaks at 630nm and 460nm which are shifted from the three experimental peaks. Or to be more explicit, the peaks of the theory in ref. 3 correspond, not with the experimental peaks, but with the minima. Fig.5b shows comparison for the experiments in ref.2 for P=600nm, $d$=250nm and $t$=340nm and again the agreement is remarkable in the peak positions, the intensities and the enhancement with respect to a single hole intensities. Our calculations are for an infinity array of holes, while in the experiment the hole arrays are finite. However the experiments [2] showed that arrays of N×N holes yield practically the same results for N>9, so the infinite arrays give, as shown by the calculations, practically the same answer at normal incidence.

We would like to explain a little bit on the appearance of the peaks positions in the periodical arrays as following. Once the cylindrical surface plasmons are excited, there will be a comparative large transmission per hole. These plasmons radiate waves at the surface and then interfere. The peaks positions and their intensity are given by the value of the period P. The hole´s diameter intervenes in the peak intensities because when, for a given frequency, a large value of a single hole intensity falls at the same position that the ideal interference peak, then this interference peak shows a pronounced maximum. However if these conditions do not match, the peak of the array is much smaller. As an illustration we present calculations in Fig.5c for $d$=200nm and P=600nm. It is clearly seen that the enhanced peak at around $\lambda \approx$690nm is strongly reduced (compared with Fig.5a, 5b for the same P value), because the single hole at this wavelength has little intensity as shown in Fig.4b (compared with Fig. 4a for different $d$ value). This is also in excellent agreement with the data of ref.2. To provide further information as a prediction result, we present in Fig.6a a set of calculations for the values of P=750nm, 870nm and 1050nm with same $d$=270nm and $t$=340nm. In agreement with the discussion above, the intensity peaks move according to the produced interferences. This shifts their peak wavelengths with periodic parameter P. Another series of transmission are also calculated (Fig.6b) fixing P=1200nm and $t$=340nm but for different $d$= 270nm, 300nm and 360nm. This time, the peak structure is always at the same position because P is fixed. However, the peaks change their intensity because $d$ is varied. Therefore P and $d$ determined the peak position and intensities respectively. Moreover, the thickness $t$ also counts, because the material has absorption and the plasmons have certain decay length. Actually, the propagation length, the plasmons speed and the retardations, etc [15] all play a role and show up in the experiments.

**II.C. The W Case for Single Holes**

In the frequency region we discussed, Ag is considered as a special case because of its ability to support surface wave for extra transmission with respect to ideal metal. We would like to see how the holes in different real metals transmit wave from the same hole´s structure. For W, in the whole visible frequency range, the real part of the permittivity is positive and approximately constant. It behaves as a dielectric. The SPP waves cannot be supported by this metal. Let us be no so definitive because this may need further discussion. But le us accept, at least, that the SPP waves, as those existing in Ag owning to permittivity (Equ.1) cannot be hold at the surface. Then, the question is should the transmission of the holes in W be very different from Ag case?

To study the case of W in the calculation, its frequency dependent permittivity is set as:

$$\varepsilon(\omega) = \varepsilon_r + \frac{\sigma}{i\omega\varepsilon_0} \qquad (3)$$

a model used for a conductor, with $\varepsilon_r$ >0 as the real part of the permittivity and $\sigma$ as the conductivity. In the sense of optical transmission, the material governed by this model behaves as a dielectric but with big attenuation. It means that the wave will penetrate the material and meanwhile loss the energy because of the attenuation. It should be noticed that in an ideal metal there is no attenuation and neither penetration, which actually plays an important role in transmission from the hole. The experimental permittivity data [9] is shown in Fig.7, together with the fitted data by Equ.3 with $\varepsilon_r$ =4 and $\sigma$=6.46×10$^5$s/m.

The calculation result of transmission from a single hole ($d$=300nm, $t$=400nm, the profile of the holes used in ref.2, Fig.3a for arrays) in W is presented in Fig. 8, together with the transmission for the same hole in an ideal metal. Their transmission profiles are similar but very different from Ag case in Fig. 2b and 4. The strong extra transmission beyond the cutoff in Ag case does not exist, which is consistent with fact that there are not surface cylindrical plasmons for W in the large wavelengths to transmit the wave. The other fact should be noticed is that the transmission is smaller than that of an ideal metal for λ<700nm. It is understandable, as we mentioned above, the wave in the hole will constantly penetrate into the metal, and the energy will be constantly killed because of the attenuation nature of W. We verified with more calculations the transmission becomes smaller and smaller till totally dies for bigger thickness of W film. However for λ>700nm the transmittivity for W overpass that of the ideal metal. This is due to penetration, which makes the effective hole size bigger. Therefore losses and penetration of the wave interplay during the transmission.

On the other hand, there is similarity between the W case and the ideal metal case. Once we calculate for a metal with the same model in Equ.3, but increasing $\sigma$ by 200 times, the resulted transmission is almost identical as ideal metal case. To understand this result we should turn to the complex reflect index $\mathbf{n}=n+ik$, where the imaginary part of the permittivity makes $\varepsilon_i \gg \varepsilon_r$, so that $n \approx k \gg 1$. The wave's decay length of the penetration into the metal is therefore greatly reduced and the reflectivity is ~1. That is the reason this model gives the same transmission as ideal metal does. In principle, a hole in W film transmit light can be treated the same as waveguide but with large attenuation. All this is a qualitative discussion. However for fixed $d$ and $t$ of the order of the experiments, ideal waveguides is not well defined. Then ideal metal and W provide more transmission than expected by long ideal wavelengths. This needs further more discussion and the case of $|\varepsilon| \to \infty$ may be an interesting one for the $t$ values discussed in this paper, yet we do not want to conclude it.

**II.D. The W case for arrays of holes**

We also perform calculations for the transmission of hole arrays in W film. The result is compared with the experiment in ref. 2 for the same structure, $d$=300nm, $t$=400nm, P=600nm. As shown in Fig.9, the calculation fits well to the experiments, for the whole profile, the peak positions and intensities. The peak position at λ~700nm is the same as those in Fig.5a and Fig 5b, with the same periodical parameter P, but with totally different material, hole size and thickness. This strongly confirm the regularity we discussed above in Ag case, that the transmission peaks of hole arrays are determined only by arrays periodic, the same conclusion that is mentioned in ref. 2. However, the intensity of the peaks in W case is 6 times smaller than those of Ag. It is merely because the transmittivity of single hole in W is smaller, by a similar factor, than that of Ag case in Fig.2b.

**Discussions and Conclusions**

From the calculations and observations made above we reach the following conclusions:

1. One obvious point, yet many people overlooked, is that one has to use the experimental data of the dielectric properties of the metal in theoretical consideration. As well as the parameters ($d$ and $t$) of the hole used.

2. Because of the previous point, the analysis of enhancements in terms of ideal calculations with approximations, like that used with Bethe theory or ideal metal long waveguides, cannot be used because it produces mislead conclusions even if the experiments are interesting and right.

3. Plasmonic metal, in particular the paradigmatic Ag, has long cutoff wavelength in the transmission because of cylindrical surface plasmons as discussed above and in earlier references 5-7. So far, the $t$ values discussed in the experiments are shorter than the decay length of the plasmons. For larger values of $t$, the transmission is drastically reduced and only remains the waveguide modes with losses. More experiments should be performed with longer $t$ to clear up this point. Moreover we have made predictions of what may happen.

4. Dielectric metals, as W in the visible, with the permittivity as given in Fig.7 behave in a particular way. There is absorption and penetration of the wave function in the metal which reduced the transmission for short wavelengths, but increased it at large wavelength with respect to ideal metal (see Fig.8).

5. If the SPPs in the Ag cases we discussed, have an influence in the transmission of the holes arrays, it seems to be marginal; i.e. is a small factor. The experiments prove this as well as our calculations.

6. The transmission peak positions of the arrays are given by the period P, and are material independent. Their intensity amplitude is large only if the single hole transmission is large. A paradigmatic effect of this is given in Fig 5a, 5c and Fig.6.

7. It appears that for the frequency discussed, Ag holds SPPs and W does not. However the existence of other possible surface waves that could enhance the field at the surface it is not clear yet.

Insisting in the SPPs enhancements and in the marginal role that they play in the experiments at hand, we believe that it is because the structure of the holes has many large Fourier components which reduce the enhancement (see ref.11-13) of the field amplitude at the surface. However there are surface profiles of the hole´s structures that may produce the desired effect. If this is possible there could be a multiplicative effect: one due to SPPs and the other to the cylindrical waves. In this sense, to work out with Ag, W and Cr choosing carefully the structures and the frequencies may give surprises. More experimental data are also needed changing the structures and the values of $d$ and $t$. For example, what happens for squared holes?

**Acknowledgements**: We thank the European EU-FP6 Project Molecular Imaging LSHG-CT-2003-503259 for support.

**Figure Captions**

**Figure 1** Schematics of cylindrical hole structures in metal film from optical transmission calculations, (a) single hole with diameter $d$ and thickness $t$ (b) arrays of holes with periodic P.

**Figure 2a** The experimental Ag permittivity dispersion data (dots) [8] and the fitting (lines) by Drude dispersion mode in Equ. 1. $\varepsilon_1$ is the real part of the permittivity, $\varepsilon_2$ is the imaginary part of the permittivity.

**Figure 2b** Transmission coefficient of single holes in different kind of film. The diameter of the hole is $d$=270nm. For ideal metal, two cases with $t$=340nm and $t$=750nm are compared. For the Ag case, $t$=340nm. And the thin line is calculated from ideal long waveguide theory.

**Figure 3a** Calculated dispersion relations $\lambda_{sp}$ by Equ.2 for the SPPs waves in the surface of Ag cylinder with a given cylindricality $\alpha$, determined by $d$=270nm and Ag bulk plasmons $\lambda_p$=325nm. Index $n$ represents the possible SPPs mode number in the cylindrical surface. The cross points between the straight photon line and the SPPs dispersion lines indicate the possible SPPs wave modes that can be excited. For large index $n$, the dispersion lines tends to the same line.

**Figure 3b** Transmission coefficient of single holes in Ag film with fixed diameter ($d$=270nm) with different thickness $t$=340nm, 525nm and 735nm.

**Figure 4** Transmission coefficient of single holes in Ag film ($t$=340nm).
**(a)** Transmission of single hole with $d$=250nm (solid line), $d$=300nm (dash dot line) by experiment [2], with $d$=250nm (-▌-), $d$=270nm (-●-) by FDTD calculation and by ideal metal waveguide theory (thick line). The enhancement factor (dash line) is presented by dividing transmission from FDTD calculation by that from waveguide theory. The arrows indicate the positions for different cylindrical plasmons excitation modes. Notice the similar oscillations appear in the experiment data with a little shift.
**(b)** Transmission of holes with $d$=200nm by experiment (solid line) [2], by FDTD calculation (-●-), by the ideal metal waveguide theory (thick line). The enhancement factor (dash line) is presented by dividing transmission from FDTD calculation by that from waveguide theory.

**Figure 5** Transmission coefficient of periodic arrays (P=600nm) of holes in Ag film

**(a)** Transmission of holes ($d$=270nm, $t$=225nm) by experiments (solid line) and by theoretical calculations in Fig. 1 of ref. 3 (dash line). Line (-•-) is our FDTD calculation results in excellent agreement with experiments.

**(b)** Transmission of holes ($d$=250nm, $t$=340nm) by the experiments (solid line) and the corresponding enhancement factor (dash line) versus single hole transmission (Fig. 2a, 2b in ref. 2). Line (-•-) is by our FDTD calculation, together with its corresponding enhancement factor versus single hole transmission (dot line).

**(c)** Transmission of holes ($d$=250nm, $t$=340nm) by the experiments (dash line) (Fig. 2c in ref. 2) and by our FDTD calculations (solid line).

**Figure 6** Transmission coefficient of periodic arrays of holes in Ag film ($t$=340nm)

**(a)** Transmission of holes with fixed $d$=270nm but for different periodic P=750nm, 870nm and 1050nm. The comparison shows the peak positions is strictly corresponding to the P values.

**(b)** Transmission of holes with fixed P=1200nm but for different diameter holes with $d$=270nm, $d$=300nm and 360nm. The comparison shows the peak positions remain fixed because of the same P, the peak intensity is influenced by the hole diameter.

**Figure 7** The experimental W permittivity dispersion data (points) [9] and the fitting (lines) by dispersion relation Equ. 3. $\varepsilon_i$ is the real part of the permittivity, $\varepsilon_r$ is the imaginary part of the permittivity.

**Figure 8** Transmission coefficient of single holes in different kind of film. The transmission of single hole in W with $d$=300nm and $t$=400nm is compared with that of the same hole in ideal metal.

**Figure 9** Transmission coefficient of periodic arrays of holes in W film with $d$=200nm, $t$=340nm and P=600nm. Experimental data (solid line) and calculation (-•-) are compared, showing well agreement.

# Figures

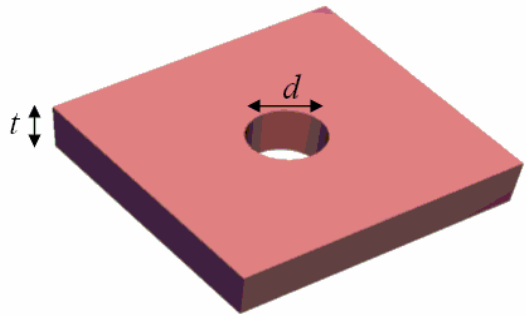 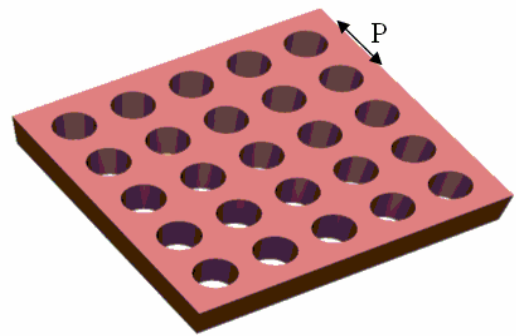

**Figure 1a**  **Figure 1b**

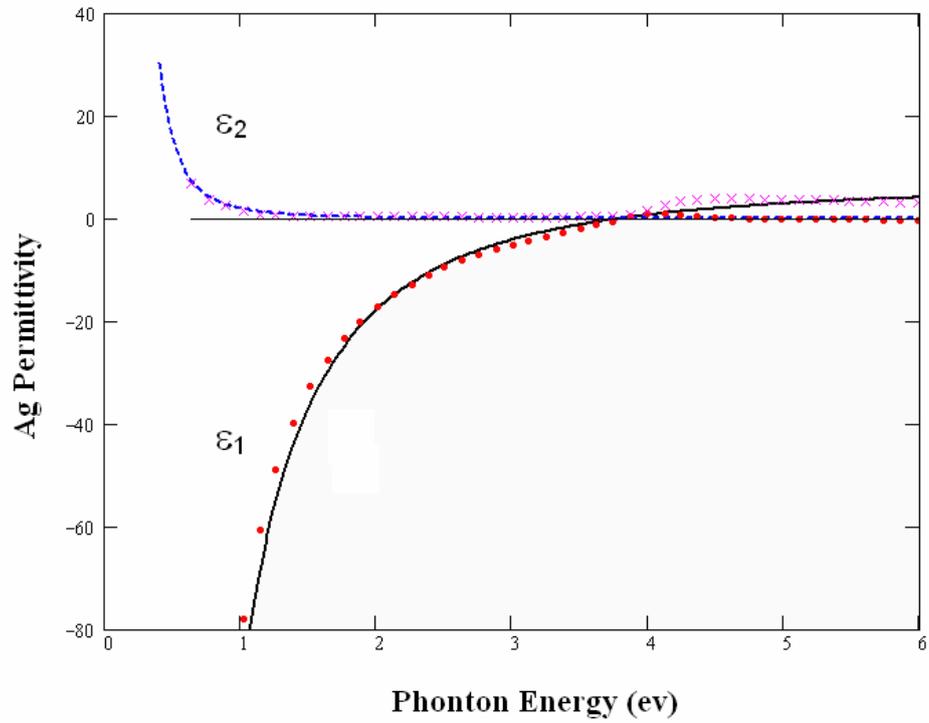

**Figure 2a**

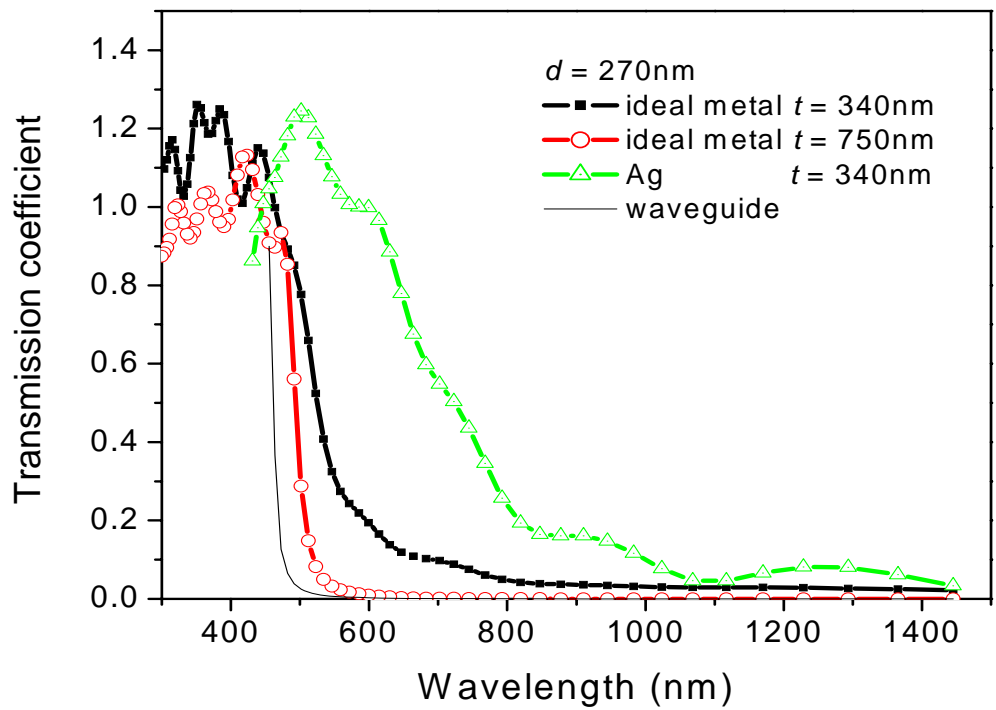

**Figure 2b**

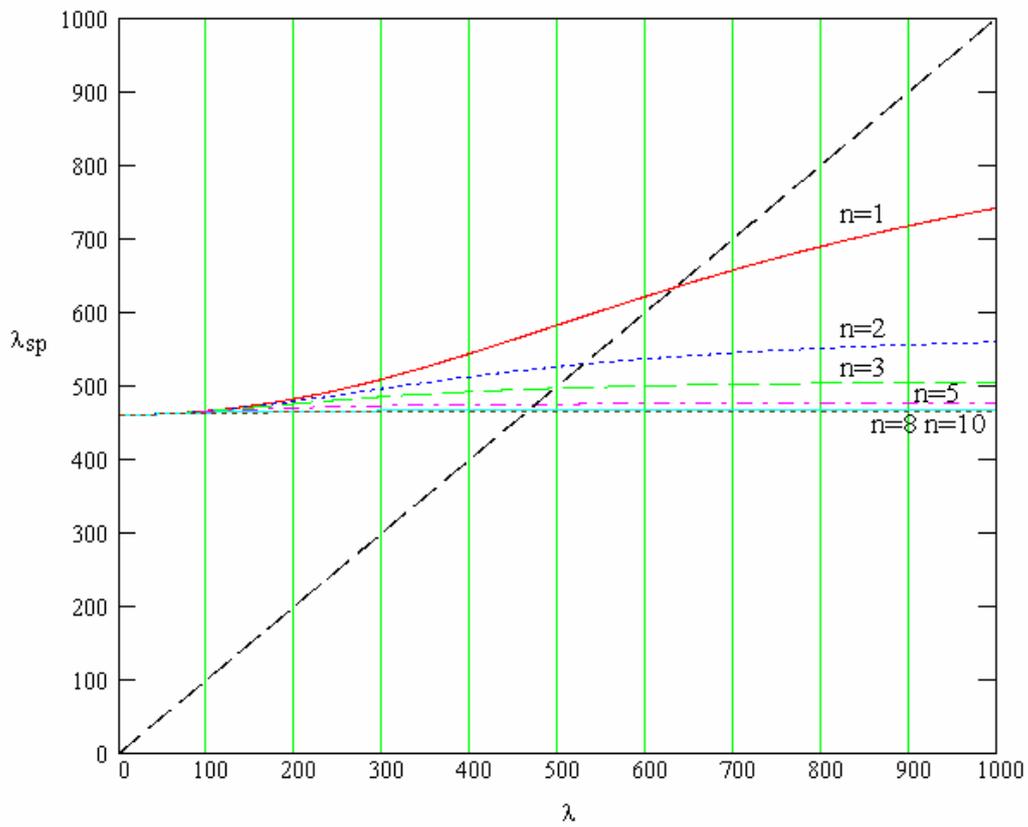

**Figure 3a**

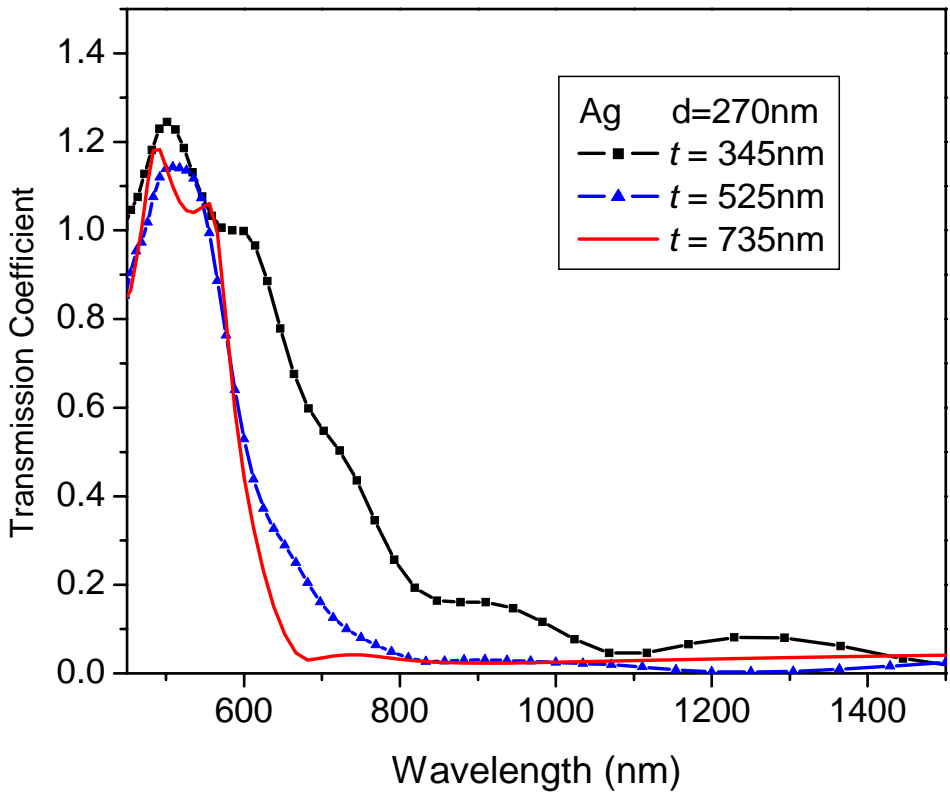

**Figure 3b**

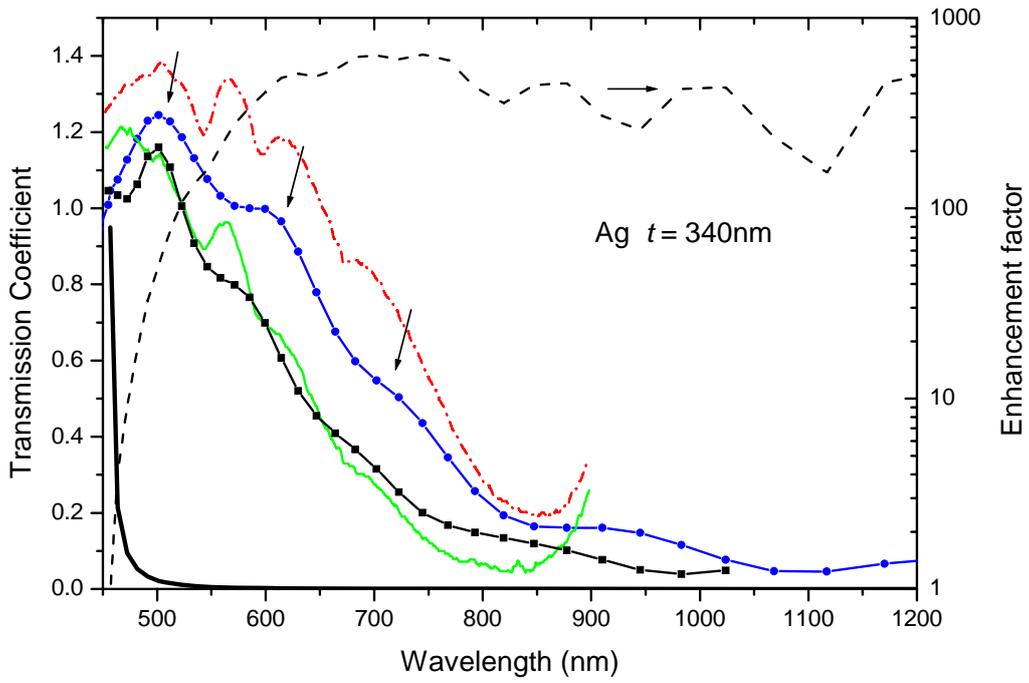

**Figure 4a**

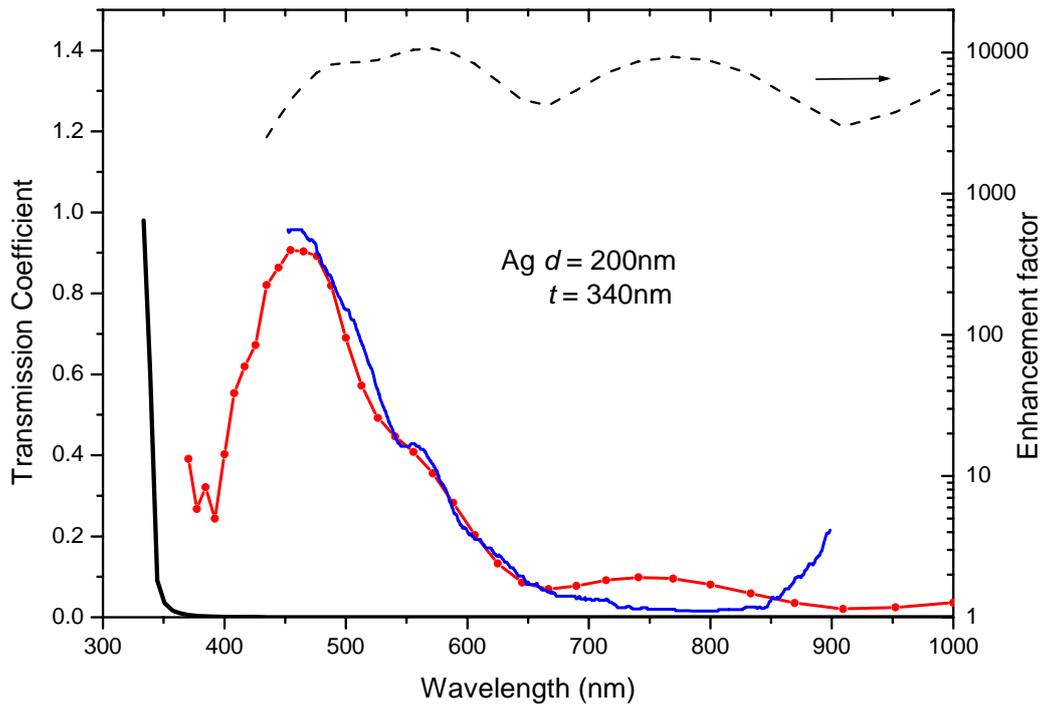

**Figure 4b**

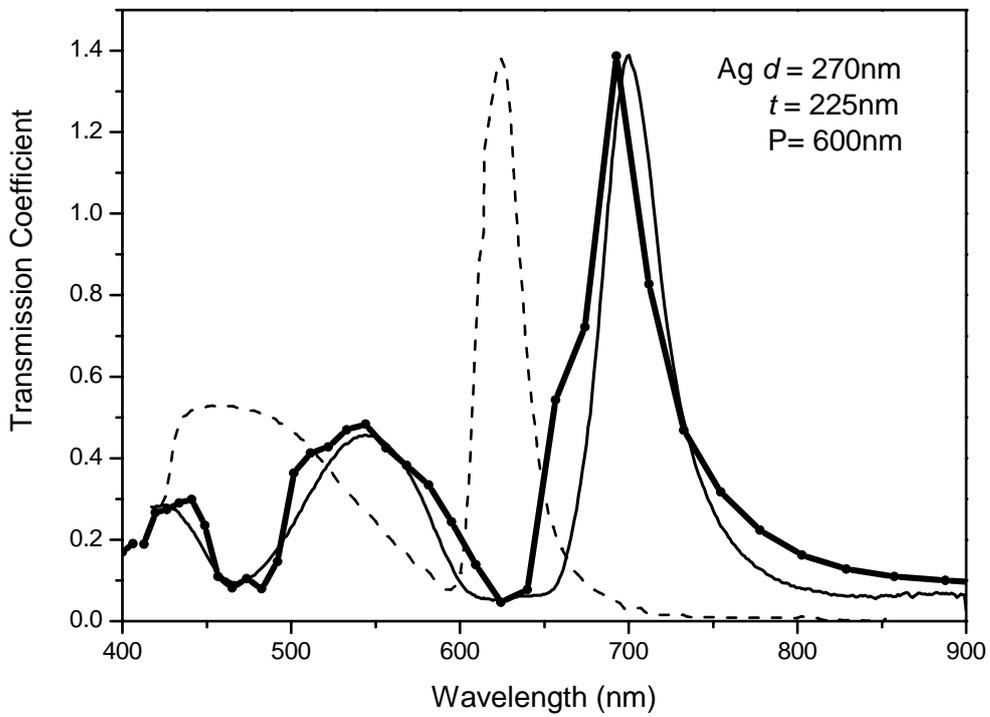

**Figure 5a**

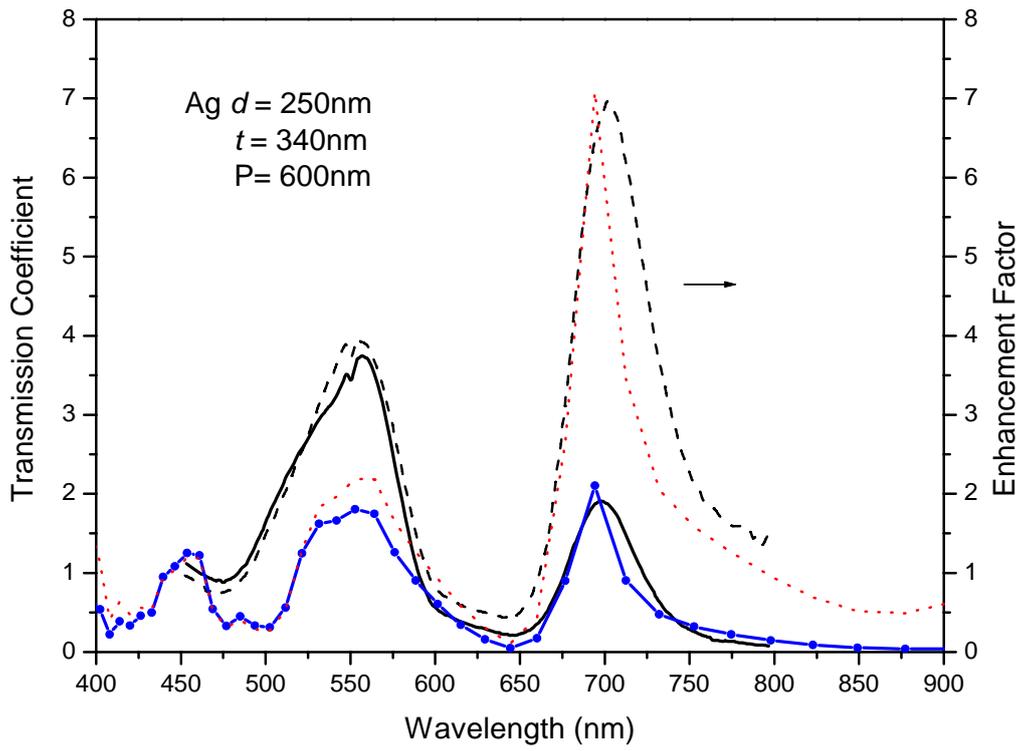

**Figure 5b**

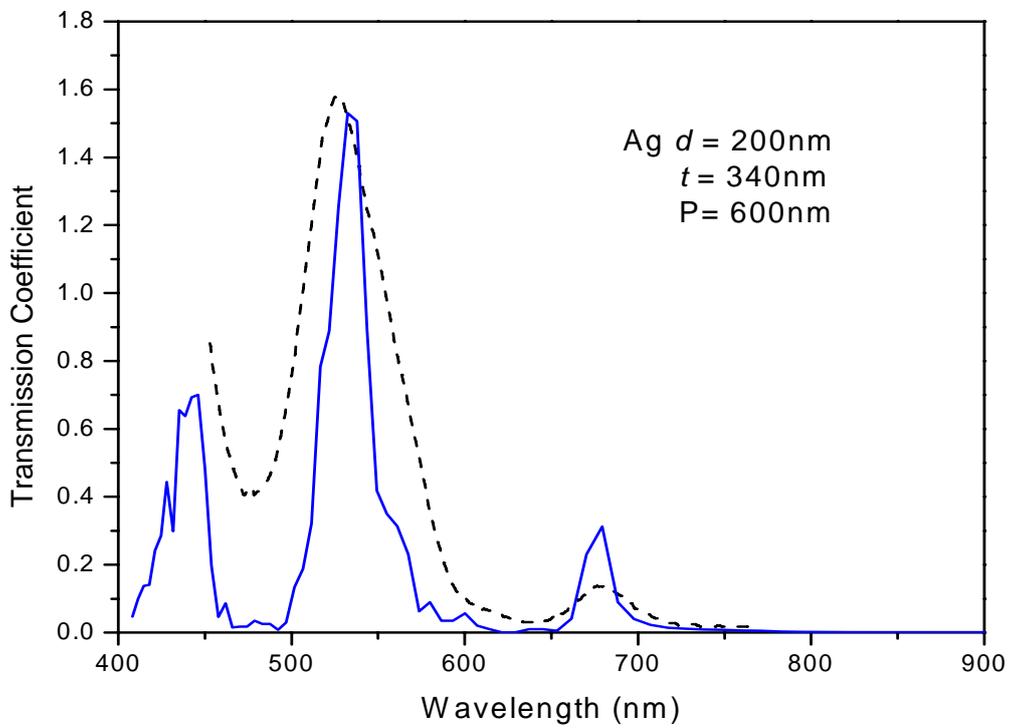

**Figure 5c**

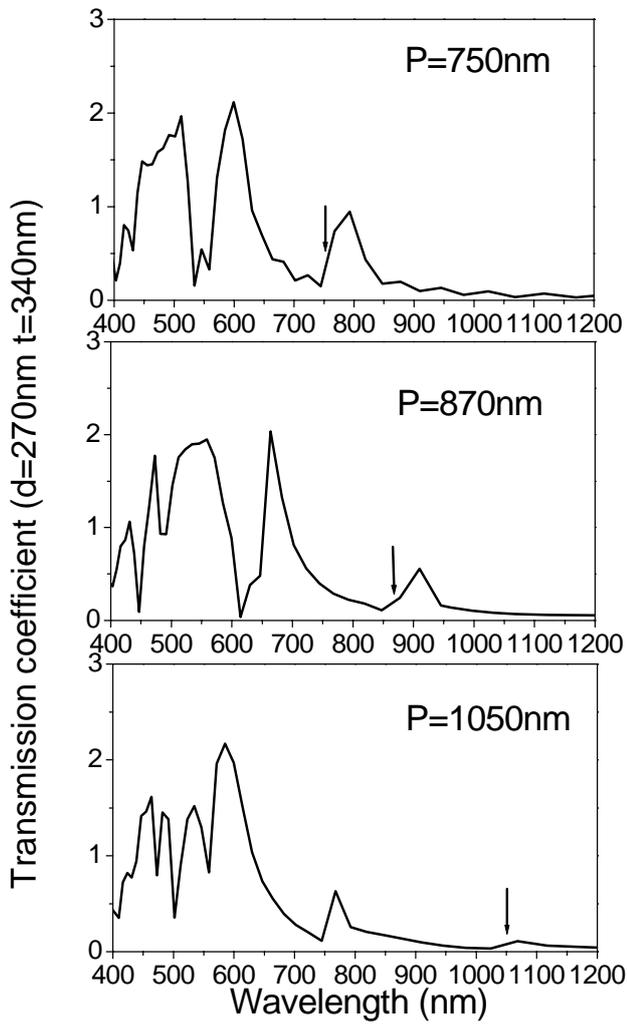 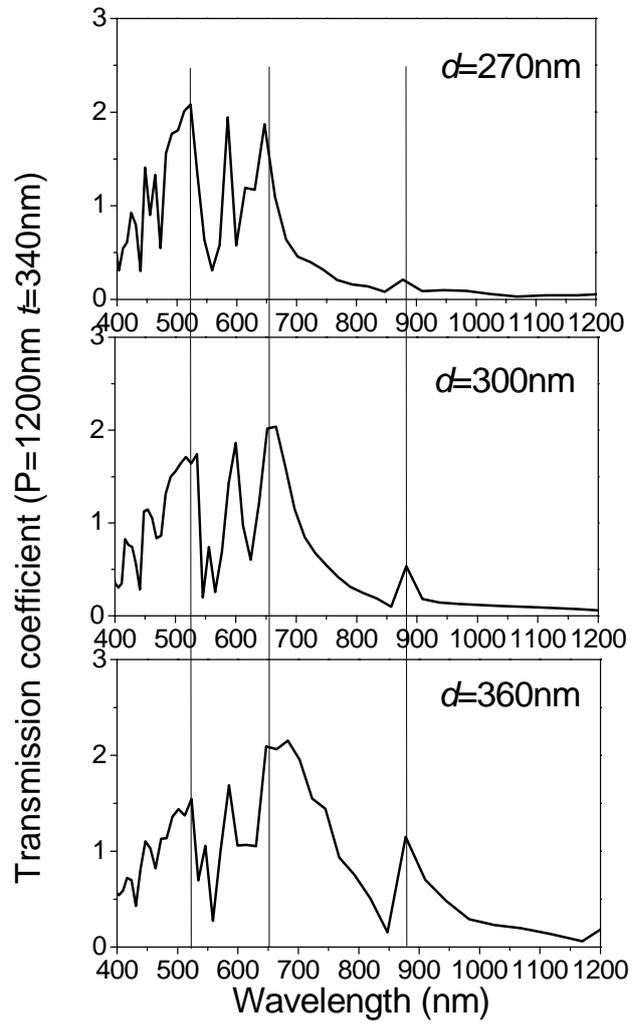

**Figure 6a**            **Figure 6b**

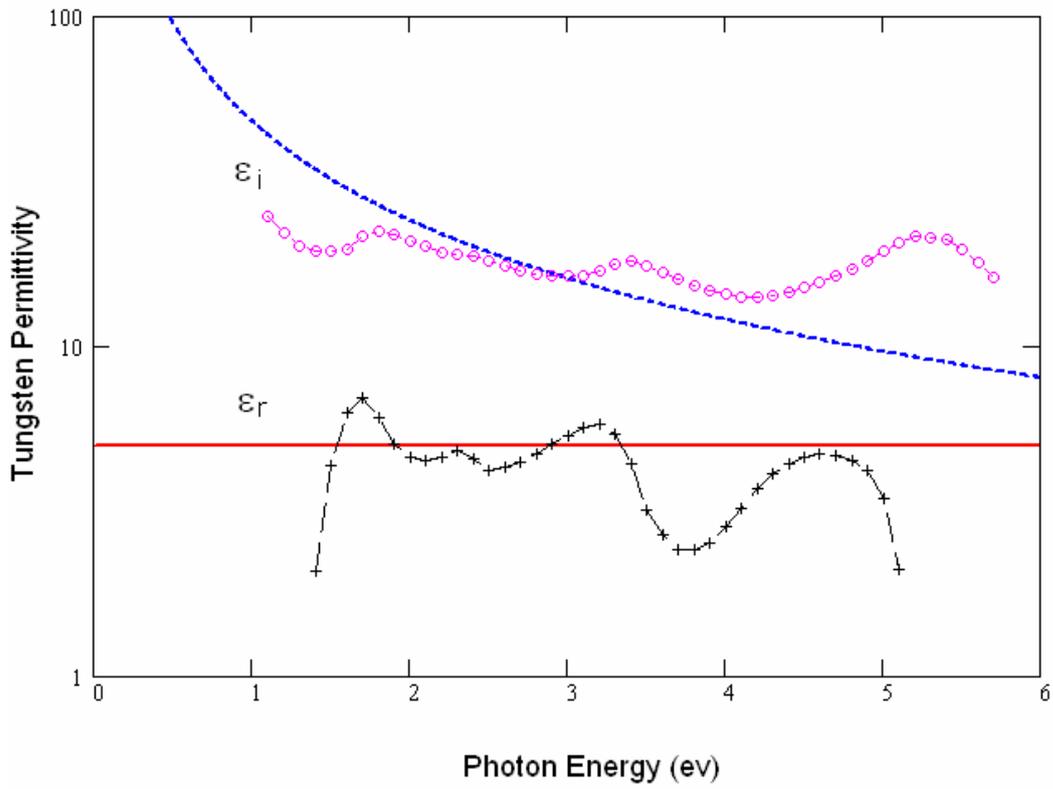

**Figure 7**

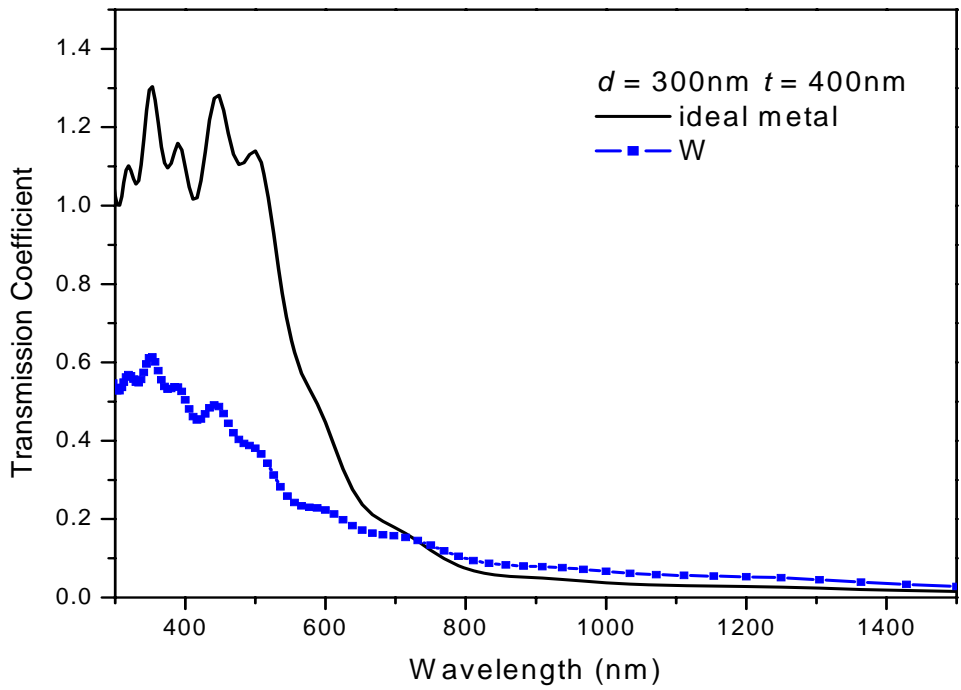

**Figure 8**

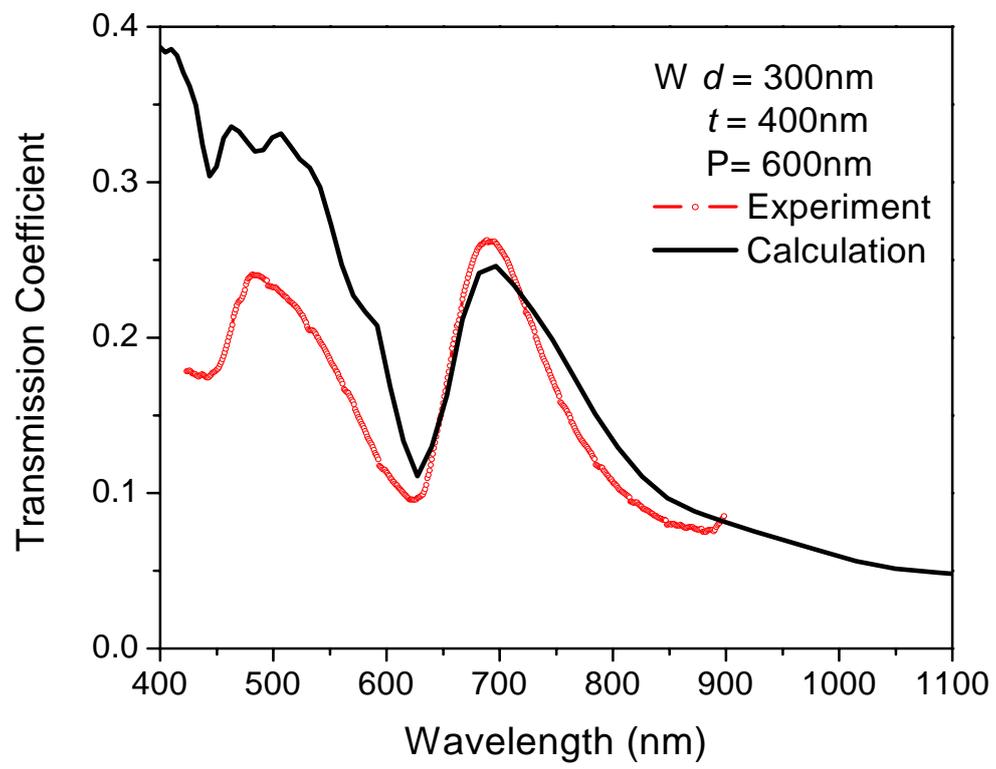

**Figure 9**